\documentclass[twocolumn,showpacs,preprintnumbers,amsmath,amssymb]{revtex4}

\usepackage{graphicx}
\usepackage{times}
\begin{document}

\title{Parton-hadron dynamics in heavy-ion collisions}

\author{E.~L.~Bratkovskaya$^{1,2}$, V. Ozvenchuk$^{2,3}$, W. Cassing$^4$,
V.~P.~Konchakovski$^4$, O.~Linnyk$^4$, R.~Marty$^{1,2}$, and H.~Berrehrah$^{1,2}$}

\affiliation{
$^{1}$ Institute for Theoretical Physics, University of Frankfurt, Frankfurt,
 Germany }
\affiliation{$^{2}$  Frankfurt Institute for Advanced Study, Frankfurt am Main,
  Germany }
\affiliation{$^{3}$ SUBATECH, UMR 6457, Laboratoire de Physique Subatomique et des
Technologies Associ\'ees, University of Nantes - IN2P3/CNRS - Ecole des Mines de Nantes, %
Nantes Cedex 03, France }
\affiliation{$^{4}$ Institute for Theoretical Physics, University of Giessen, Giessen, Germany }

\date{\today}

\begin{abstract}
The dynamics of partons and hadrons in relativistic nucleus-nucleus
collisions is analyzed within the novel Parton-Hadron-String
Dynamics (PHSD) transport approach, which is based on a dynamical
quasiparticle model for the partonic phase (DQPM) including a
dynamical hadronization scheme. The PHSD approach is applied to
nucleus-nucleus collisions from low SPS to LHC energies. The traces
of partonic interactions are found in particular in the elliptic
flow of hadrons and in their transverse mass spectra. We investigate
also the equilibrium properties of strongly-interacting infinite
parton-hadron matter characterized by transport coefficients such as
shear and bulk viscosities and the electric conductivity in
comparison to lattice QCD results.
\end{abstract}

\maketitle

\section{Introduction}

According to present understanding our universe has been created in
a 'Big Bang' about 1,37$\cdot$10$^{10}$ years ago. The 'Big Bang'
scenario implies that in the first micro-seconds of the universe the
entire state has emerged from a partonic system of quarks,
antiquarks and gluons -- a quark-gluon plasma (QGP) -- to color
neutral hadronic matter consisting of interacting hadronic states
(and resonances) in which the partonic degrees of freedom are
confined. The nature of confinement and the dynamics of this phase
transition has motivated a large community for several decades and
is still an outstanding question of todays physics. Early concepts
of the QGP were guided by the idea of a weakly interacting system of
partons which might be described by perturbative QCD (pQCD).
However, experimental observations at the Relativistic Heavy Ion
Collider (RHIC) indicated that the new medium created in
ultrarelativistic Au+Au collisions is interacting more strongly than
hadronic matter and consequently this concept had to be severely
questioned. Moreover, in line with theoretical studies in
Refs.~\cite{Shuryak,Thoma,Andre} the medium showed phenomena of an
almost perfect liquid of partons~\cite{STARS,Miklos3} as extracted
from the strong radial expansion and the scaling of elliptic flow
$v_2(p_T)$ of mesons and baryons with the number of constituent
quarks and antiquarks~\cite{STARS}. Indeed, recent relativistic
viscous hydrodynamic calculations - using the Israel-Stewart
framework - require a very small  shear viscosity to entropy density
ratio $\eta/s$ of $0.08-0.24$ in order to reproduce the  elliptic
flow $v_2$ data at RHIC (cf. \cite{ViscousHydro1}). There is strong
evidence from atomic and molecular systems that $\eta/s$ should have
a minimum in the vicinity of the phase transition (or rapid
crossover) between the hadronic matter and the quark-gluon plasma
\cite{Minshear}, and that the ratio of bulk viscosity to entropy
density $\zeta/s$ should be maximum or even diverge at a
second-order phase transition \cite{MaxBulk1}. It is also important
to know the electromagnetic properties of the QGP, such as the
electric conductivity, since it determines the electromagnetic
radiation in terms of photons and dileptons.

The question about the properties of this (nonperturbative) QGP
'liquid' is discussed controversially in the literature and
dynamical concepts describing the formation of color neutral hadrons
from colored partons are scarce. A fundamental issue for
hadronization models is the conservation of 4-momentum as well as
the entropy problem, because by fusion/coalescence of massless (or
low constituent mass) partons to color neutral bound states of low
invariant mass (e.g. pions) the number of degrees of freedom and
thus the total entropy is reduced in the hadronization process. This
problem - a violation of the second law of thermodynamics as well as
the conservation of four-momentum and flavor currents - has been
addressed in Ref.~\cite{PRC08} on the basis of the DQPM employing
covariant transition rates for the fusion of 'massive' quarks and
antiquarks to color neutral hadronic resonances or strings. In fact,
the dynamical studies for an expanding partonic fireball in
Ref.~\cite{PRC08} suggest that the these problems have come to a
practical solution.

A consistent dynamical approach - valid also for strongly interacting
systems - can be formulated on the basis of Kadanoff-Baym (KB)
equations~\cite{Sascha1} or off-shell transport equations in
phase-space representation,
respectively~\cite{Sascha1}. In the KB theory the field
quanta are described in terms of dressed propagators with complex
selfenergies.  Whereas the real part of the selfenergies can be
related to mean-field potentials (of Lorentz scalar, vector or tensor
type), the imaginary parts provide information about the lifetime
and/or reaction rates of time-like 'particles'~\cite{Crev}. Once the
proper (complex) selfenergies of the degrees of freedom are known the
time evolution of the system is fully governed by off-shell transport
equations (as described in Refs.~\cite{Sascha1,Crev}).  The
determination/extraction of complex selfenergies for the partonic
degrees of freedom has been performed before in
Ref.~\cite{Cassing07} by fitting lattice QCD (lQCD) 'data'
within the Dynamical QuasiParticle Model (DQPM). In fact, the DQPM
allows for a simple and transparent interpretation of lattice QCD
results for thermodynamic quantities as well as correlators and leads
to effective strongly interacting partonic quasiparticles with broad
spectral functions.  For a review on off-shell transport theory and
results from the DQPM in comparison to lQCD we refer the reader to
Ref.~\cite{Crev}.

The actual implementations in the PHSD transport approach have been
presented in detail in Refs.~\cite{PHSD,BCKL11}. Here we  present
results for transverse mass spectra and elliptic flow of hadrons for
heavy-ion collisions at relativistic energies in comparison to data
from the experimental collaborations.

\section{The PHSD approach}

The dynamics of partons, hadrons and strings in relativistic
nucleus-nucleus collisions is analyzed here within the
Parton-Hadron-String Dynamics approach~\cite{PRC08,PHSD,BCKL11}. In this
transport approach the partonic dynamics is based on Kadanoff-Baym
equations for Green functions with self-energies from the Dynamical
QuasiParticle Model (DQPM) \cite{Cassing07} which
describes QCD properties in terms of 'resummed' single-particle
Green functions. In Ref.~\cite{BCKL11}, the actual  three DQPM
parameters for the temperature-dependent effective coupling were
fitted to the recent lattice QCD results of Ref.~\cite{aori10}.
The latter lead to a critical temperature $T_c \approx$ 160 MeV
which corresponds to a critical energy density of $\epsilon_c
\approx$ 0.5 GeV/fm$^3$. In PHSD the parton spectral functions
$\rho_j$ ($j=q, {\bar q}, g$) are no longer $\delta-$ functions in
the invariant mass squared as in conventional cascade or transport
models but depend on the parton mass and width parameters:
\begin{eqnarray}
\!\!\!\!\!\! \rho_j(\omega,{\bf p}) =
 \frac{\gamma_j}{E_j} \left(
   \frac{1}{(\omega-E_j)^2+\gamma_j^2} - \frac{1}{(\omega+E_j)^2+\gamma_j^2}
 \right)\
\label{eq:rho}
\end{eqnarray}
separately for quarks/antiquarks and gluons ($j=q,\bar{q},g$).
With the convention $E^2({\bf p}^2) = {\bf p}^2+M_j^2-\gamma_j^2$, the
parameters $M_j^2$ and $\gamma_j$ are directly related to the real
and imaginary parts of the retarded self-energy, {\it e.g.} $\Pi_j =
M_j^2-2i\gamma_j\omega$. The spectral function~(\ref{eq:rho}) is
antisymmetric in $\omega$ and normalized as
\begin{equation}
 \int_{-\infty}^{\infty} \frac{d \omega}{2 \pi} \
 \omega \ \rho_j(\omega, {\bf p}) = \int_0^{\infty} \frac{d
 \omega}{2 \pi} \ 2 \omega \ \rho_j(\omega, {\bf p}) = 1 \ .
\label{normalize}
\end{equation}

\begin{figure}[tbh]
 \includegraphics[width=70mm]{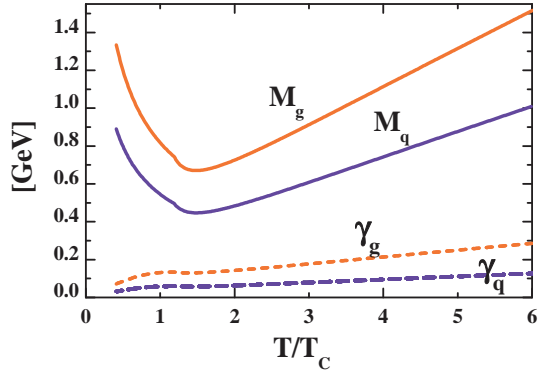}
\caption{ The effective gluon mass
$M_g$ and witdh $\gamma_g$ as function of the scaled temperature $T/T_c$ (red lines).
The blue lines show the corresponding quantities for quarks.}
\label{fig1}
\end{figure}

The actual parameters in Eq.~(\ref{eq:rho}), {\it i.e.} the gluon mass
$M_g$ and width $\gamma_g$ -- employed as input in the PHSD
calculations -- as well as the quark mass $M_q$ and width
$\gamma_q$, are depicted in Fig. \ref{fig1} as a
function of the scaled temperature $T/T_c$. As mentioned above these
values for the masses and widths have been fixed by fitting the
lattice QCD results from Ref.~\cite{aori10} in thermodynamic
equilibrium.

\begin{figure}[tb]
\centering \includegraphics*[width=70mm]{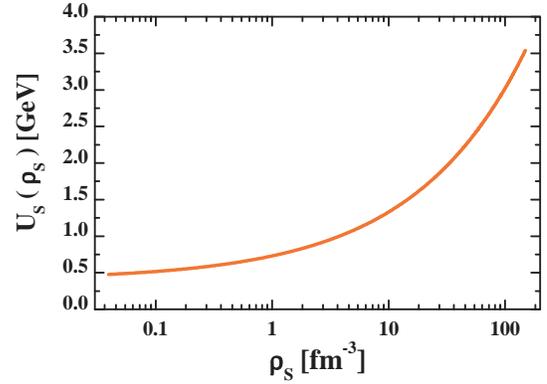} \caption{ The
scalar mean field (\ref{uss}) for quarks and antiquarks from the
DQPM as a function of the scalar parton density $\rho_s$ (2).}
\label{figpot}
\end{figure}

A scalar mean-field $U_s(\rho_s)$ for quarks and antiquarks can be defined by the
derivative,
\begin{equation} \label{uss}
U_s(\rho_s) = \frac{d V_p(\rho_s)}{d \rho_s} ,
\end{equation}
which is evaluated numerically within the DQPM.
Here $V_p$ is a  a potential energy density
\begin{equation} \label{Vp}
V_p(T,\mu_q) = T^{00}_{g-}(T,\mu_q) + T^{00}_{q-}(T,\mu_q) + T^{00}_{{\bar q}-}(T,\mu_q)
\end{equation}
where the different contributions $T^{00}_{j-}$ correspond to the space-like part
of the energy-momentum tensor component $T^{00}_{j}$ of parton $j
= g, q, \bar{q}$ (cf. Section 3 in Ref. \cite{Cassing07}).
The scalar mean-field $U_s(\rho_s)$ for quarks and antiquarks is displayed in Fig. \ref{figpot}
as a function of the parton scalar density $\rho_s$ and shows that
the scalar mean field is in the order of a few GeV for $\rho_s >
10$ fm$^{-3}$. The mean-field (\ref{uss}) is employed in the PHSD
transport calculations and determines the force on a quasiparticle
$j$, i.e.
$ \sim M_j/E_j \nabla U_s(x) = M_j/E_j \
d U_s/d \rho_ s \ \nabla \rho_s(x)$ where the scalar density
$\rho_s(x)$ is determined numerically on a space-time grid.

Furthermore, a two-body
interaction strength can be extracted from the DQPM as well from the
quasiparticle width in line with Ref.~\cite{Andre}. The transition
from partonic to hadronic d.o.f. (and vice versa)  is
described by covariant transition rates for the fusion of
quark-antiquark pairs or three quarks (antiquarks), respectively,
obeying flavor current-conservation, color neutrality as well as
energy-momentum conservation~\cite{PHSD,BCKL11}. Since the dynamical
quarks and antiquarks become very massive close to the phase
transition, the formed resonant 'prehadronic' color-dipole states
($q\bar{q}$ or $qqq$) are of high invariant mass, too, and
sequentially decay to the groundstate meson and baryon octets
increasing the total entropy.

On the hadronic side PHSD includes explicitly the baryon octet and
decouplet, the $0^-$- and $1^-$-meson nonets as well as selected
higher resonances as in the Hadron-String-Dynamics (HSD)
approach~\cite{Ehehalt,HSD,Cass02}. The color-neutral objects of
higher masses ($>$1.5~GeV in case of baryonic states and $>$1.3~GeV
in case of mesonic states) are treated as `strings' (color-dipoles)
that decay to the known (low-mass) hadrons according to the JETSET
algorithm~\cite{JETSET}. We discard an explicit recapitulation of
the string formation and decay and refer the reader to the original
work~\cite{JETSET}. Note that PHSD and HSD (without explicit
partonic degrees-of-freedom) merge at low energy density, in
particular below the critical energy density $\varepsilon_c\approx$
0.5~GeV/fm$^{3}$.

\section{Application to nucleus-nucleus collisions}

The PHSD approach was applied to nucleus-nucleus collisions from
$s_{NN}^{1/2} \sim$ 5 to 200 GeV in Refs.~\cite{PHSD,BCKL11,Voka12} in order
to explore the space-time regions of 'partonic matter'. It was found
that even central collisions at the top-SPS energy of
$\sqrt{s_{NN}}=$17.3 GeV show a large fraction of nonpartonic, {\it
i.e.} hadronic or string-like matter, which can be viewed as a hadronic
corona~\cite{Aichelin}. This finding implies that neither hadronic nor
only partonic models can be employed to extract physical conclusions in
comparing model results with data.

\subsection{Transverse mass spectra}

\begin{figure*}
\phantom{a}\hspace*{-2.5cm}
  \includegraphics[width=0.65\textwidth]{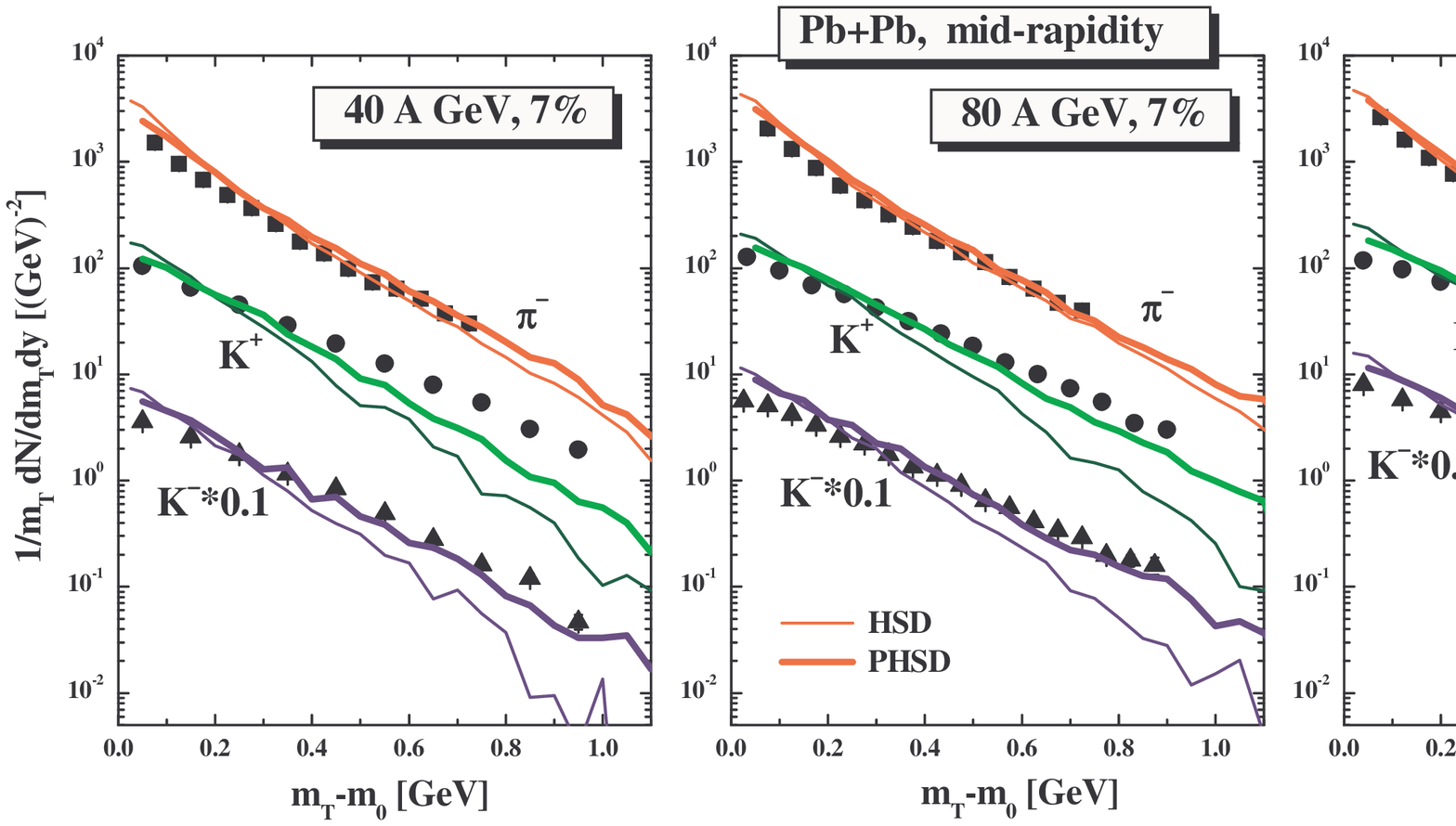}
  \caption{The $\pi^-$, $K^+$ and $K^-$ transverse mass spectra for
    central Pb+Pb collisions at 40, 80 and 158 A$\cdot$GeV from PHSD
    (thick solid lines) in comparison to the distributions from HSD
    (thin solid lines) and the experimental data from the NA49
    Collaboration~\cite{NA49a}. }
  \label{fig11}
\end{figure*}

It is of interest, how the PHSD approach compares to the
HSD~\cite{HSD} model (without explicit partonic degrees-of-freedom)
as well as to experimental data. In Fig.~\ref{fig11} we show the
transverse mass spectra of $\pi^-$, $K^+$ and $K^-$ mesons for 7\%
central Pb+Pb collisions at 40 and 80 A$\cdot$GeV and 5\% central
collisions at 158 A$\cdot$GeV in comparison to the data of the NA49
Collaboration~\cite{NA49a}.  Here the slope of the $\pi^-$ spectra
is only slightly enhanced in PHSD relative to HSD which demonstrates
that the pion transverse motion shows no sizeable sensitivity to the
partonic phase. However, the $K^\pm$ transverse mass spectra are
substantially hardened with respect to the HSD calculations at all
bombarding energies - i.e. PHSD is more in line with the data - and
thus suggests that partonic effects are better visible in the
strangeness-degrees of freedom.

\begin{figure}
\phantom{a}\vspace*{1mm}
\includegraphics[width=0.35\textwidth]{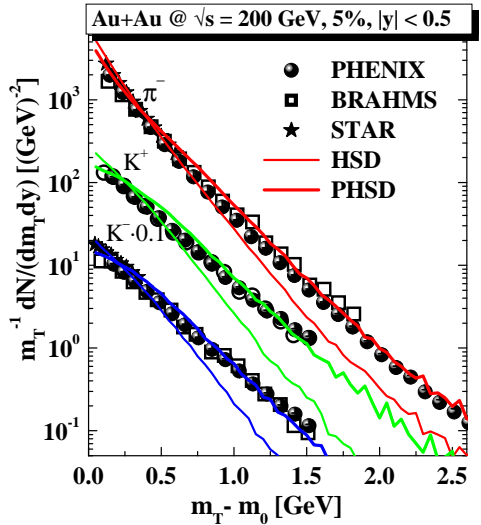}
\caption{The $\pi^-$, $K^+$ and $K^-$ transverse mass spectra for
5\%
  central Au+Au collisions at $\sqrt{s}$ = 200 GeV from PHSD (thick solid
  lines) in comparison to the distributions from HSD (thin solid
  lines) and the experimental data from the BRAHMS, PHENIX and STAR
  Collaborations~\cite{PHENIX2,STAR3,BRAHMS} at midrapidity.}
\label{Fig_mtRHIC}
\end{figure}

The PHSD calculations for RHIC energies show a very similar trend -
the inverse slope increases by including the partonic phase -
cf. Fig. \ref{Fig_mtRHIC} where we show the transverse mass spectra of
$\pi^-$, $K^+$ and $K^-$ mesons for 5\% central Au+Au collisions at
$\sqrt{s}$ = 200 GeV in comparison to the data of the RHIC
Collaborations~\cite{PHENIX2,STAR3,BRAHMS}.

The hardening of the kaon spectra can be traced back to parton-parton
scattering as well as a larger collective acceleration of the partons
in the transverse direction due to the presence of repulsive vector
fields for the partons. The enhancement of the spectral slope for
kaons and antikaons in PHSD due to collective partonic flow shows up
much clearer for the kaons due to their significantly larger mass
(relative to pions). We recall that in Refs.~\cite{BratPRL} the
underestimation of the $K^\pm$ slope by HSD (and also UrQMD) had been
suggested to be a signature for missing partonic degrees of freedom;
the present PHSD calculations support this early suggestion.

The strange antibaryon sector is of further interest since here the HSD
calculations have always underestimated the yield~\cite{Geiss}.  Our
detailed studies in Ref.~\cite{PHSD} show that the HSD and PHSD
calculations both give a reasonable description of the $\Lambda +
\Sigma^0$ yield of the NA49 Collaboration~\cite{NA49_aL09}; both models
underestimate the NA57 data~\cite{NA57} by about 30\%. An even larger
discrepancy in the data from the NA49 and NA57 Collaborations is seen
for $(\bar \Lambda + \bar \Sigma^0)/N_{wound}$; here the PHSD
calculations give results which are in between the NA49 data and the
NA57 data whereas  HSD underestimates the $(\bar \Lambda + \bar
\Sigma^0)$ midrapidity yield at all centralities.

The latter result suggests that the partonic phase does not show up
explicitly in an enhanced production of strangeness (or in particular
strange mesons and baryons) but leads to a different redistribution of
antistrange quarks between mesons and antibaryons.  In fact, as
demonstrated in Ref.~\cite{PHSD}, we find no sizeable differences
in the double strange baryons from HSD and PHSD -- in a good agreement
with the NA49 data -- but observe a large enhancement in the double
strange antibaryons for PHSD relative to HSD.

\subsection{Collective flow}
 The anisotropy  in the azimuthal angle $\psi$ is usually characterized
by the even order Fourier coefficients $v_n =\langle exp(\, \imath \,
n(\psi-\Psi_{RP}))\rangle, \
 n = 2, 4, ...$, since for a smooth angular profile the odd harmonics
become equal to zero. As noted above, $\Psi_{RP}$ is the azimuth of the
reaction plane and the brackets denote averaging over particles
and events. In particular, for the widely used second order coefficient,
denoted as an elliptic flow, we have
 \begin{equation}
 \label{eqv2}
 v_2 = \left<cos(2\psi-2\Psi_{RP})\right>=
 \left<\frac{p^2_x - p^2_y}{p^2_x + p^2_y}\right>~,
\end{equation}
where $p_x$ and $p_y$ are the $x$ and $y$ components of the particle
momenta. This coefficient can be considered as a function of
centrality, pseudo-rapidity $\eta$ and/or transverse momentum $p_T$.  We
note that the reaction plane in PHSD is given by the $(x - z)$ plane
with the $z$-axis in the beam direction.

\begin{figure}[t]
\includegraphics[scale=.3]{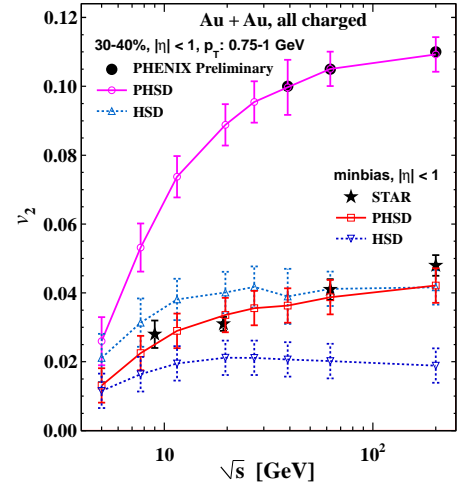}
\caption{Average elliptic flow $v_2$ of charged particles at
midrapidity for two centrality selections calculated within the PHSD
(solid curves) and HSD (dashed curves). The $v_2$ STAR data
compilation for minimal bias collisions  are taken
from~\cite{NKKNM10} (stars) and the preliminary PHENIX
data~\cite{PHENIX_v2_s} are plotted by filled circles.}
\label{vns}
\end{figure}

In Fig.~\ref{vns} the experimental $v_2$ excitation function in the
transient energy range is compared to the results from the PHSD
calculations \cite{Voka11}; HSD model results are given as well for
reference. We note that the centrality selection and acceptance are the
same for the data and models.

We recall that the  HSD model has been very successful in describing
heavy-ion spectra and rapidity distributions from SIS to SPS
energies. A detailed comparison of HSD results with respect to a
large experimental data set was  reported in
Refs.~\cite{mt-SIS,BratPRL,BRAT04}
for central Au+Au (Pb+Pb) collisions from SIS to top SPS energies.
Indeed, as shown in Fig.~\ref{vns} (dashed lines), HSD is in good
agreement with experiment for both data sets at the lower edge
($\sqrt{s_{NN}}\sim$10 GeV)  but predicts an approximately
energy-independent flow $v_2$ at  larger energies and, therefore, does
not match the experimental observations. This behavior is in quite
close agreement with another independent hadronic model, the UrQMD
(Ultra relativistic Quantum Molecular Dynamics)~\cite{UrQMD} (cf.
with Ref.~\cite{NKKNM10}).

\noindent

\begin{figure}[h]
\includegraphics[scale=.3]{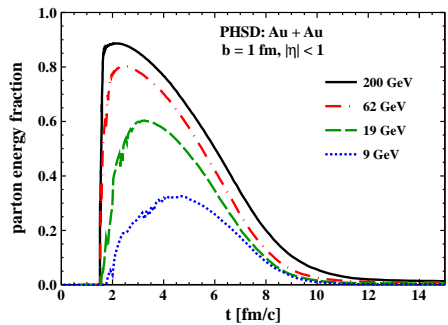}
\caption{The evolution of the parton fraction of the total energy
  density at mid-pseudorapidity for different collision energies.}
\label{part}
\end{figure}

From the above comparison one may conclude  that the rise of $v_2$
with bombarding energy is not due to hadronic interactions and
models with partonic d.o.f. have to be addressed.
Indeed, the PHSD approach incorporates the parton medium effects in
line with a lQCD equation-of-state, as discussed above, and also
includes a dynamic hadronization scheme based on covariant
transition rates. It is seen from Fig.~\ref{vns} that PHSD performs
better: The elliptic flow $v_2$ from PHSD (solid curve) is fairly in
line with the data from the STAR and PHENIX collaborations and
clearly shows the growth of $v_2$ with the bombarding energy \cite{Voka11}.

The $v_2$ increase is clarified in Fig.~\ref{part} where the partonic
fraction of the energy density at mid-pseudorapidity with respect to
the total energy density in the same pseudorapidity interval is
shown. We recall that the repulsive mean-field potential
$U(\rho_v)$ for partons in the PHSD model leads to an increase of
the flow $v_2$ as compared to that for HSD or PHSD calculations
without partonic mean fields.  As follows from Fig.~\ref{part}, the
energy fraction of the partons substantially grows with increasing
bombarding energy while the duration of the partonic phase is roughly
the same.

\section{Transport coefficients}

In this Section we concentrate on the extraction of equilibrium
properties of  'infinite' parton-hadron matter characterized by
transport coefficients such as shear and bulk viscosity and electric
conductivity which have been studied within the PHSD approach in
Refs. \cite{box,Rudy13,Cassing:2013iz}.

We simulate the `infinite' matter within a cubic box
with periodic boundary conditions at various values for the quark
density (or chemical potential) and energy density. The size of the
box is fixed to $9^3$ fm$^3$. The initialization is done by
populating the box with light ($u,d$) and strange ($s$) quarks,
antiquarks and gluons. The system is initialized out of equilibrium
and approaches kinetic and chemical equilibrium during it's
evolution by PHSD. If the energy density in the system is below the
critical energy density ($\varepsilon_c \approx $ 0.5 GeV/fm$^3$),
the evolution proceeds through the dynamical phase transition (as
described in Ref. \cite{box}) and ends up in an ensemble
of interacting hadrons.
The transport properties are calculated employing the
Kubo formalism \cite{Green,Kubo} and relaxation time approximation
\cite{Hosoya,Gavin,Kapusta}.
For more details we refer the reader to Ref.~\cite{box}.

\begin{figure*}
\centering
\includegraphics[width=0.32\textwidth,clip]{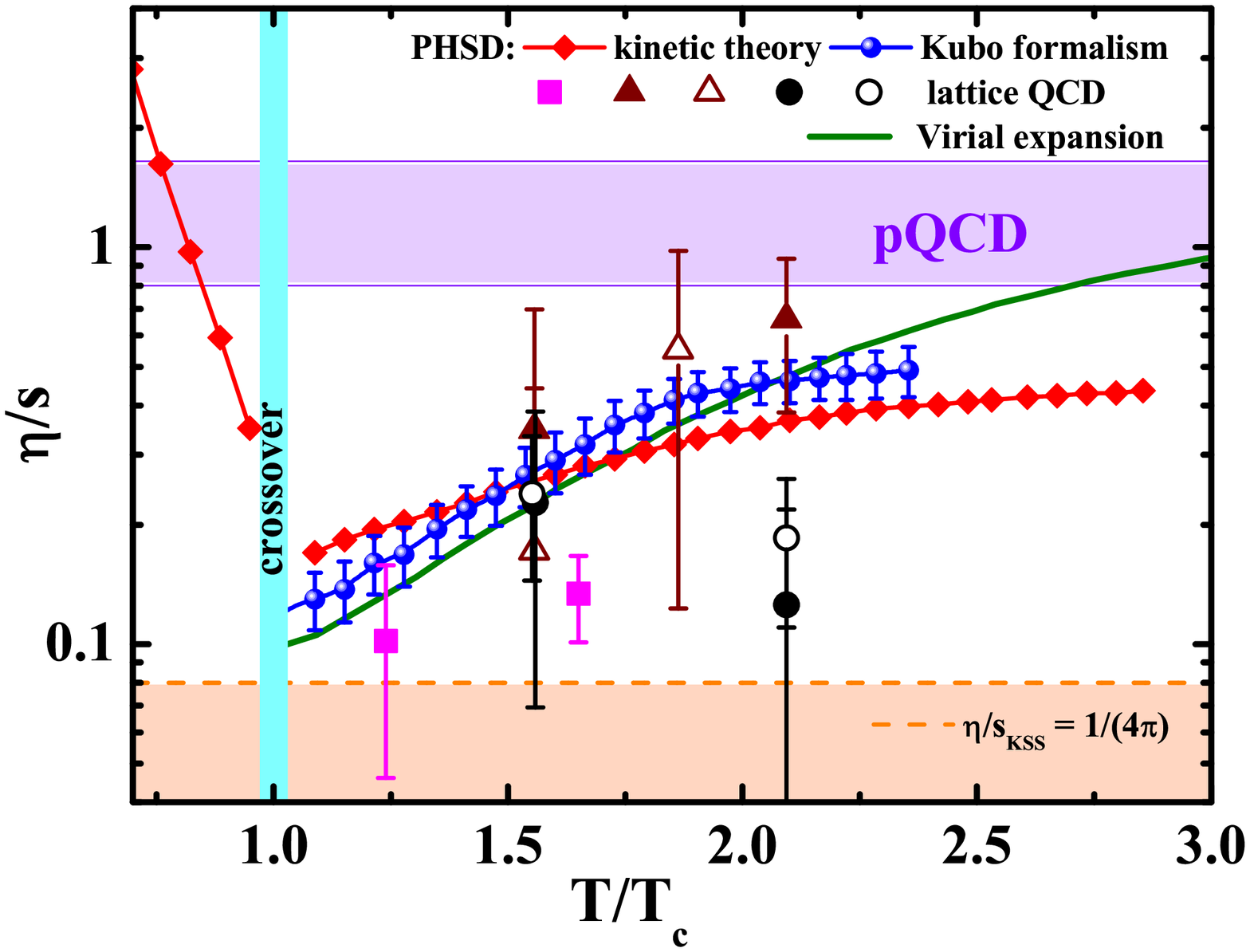}
\includegraphics[width=0.32\textwidth,clip]{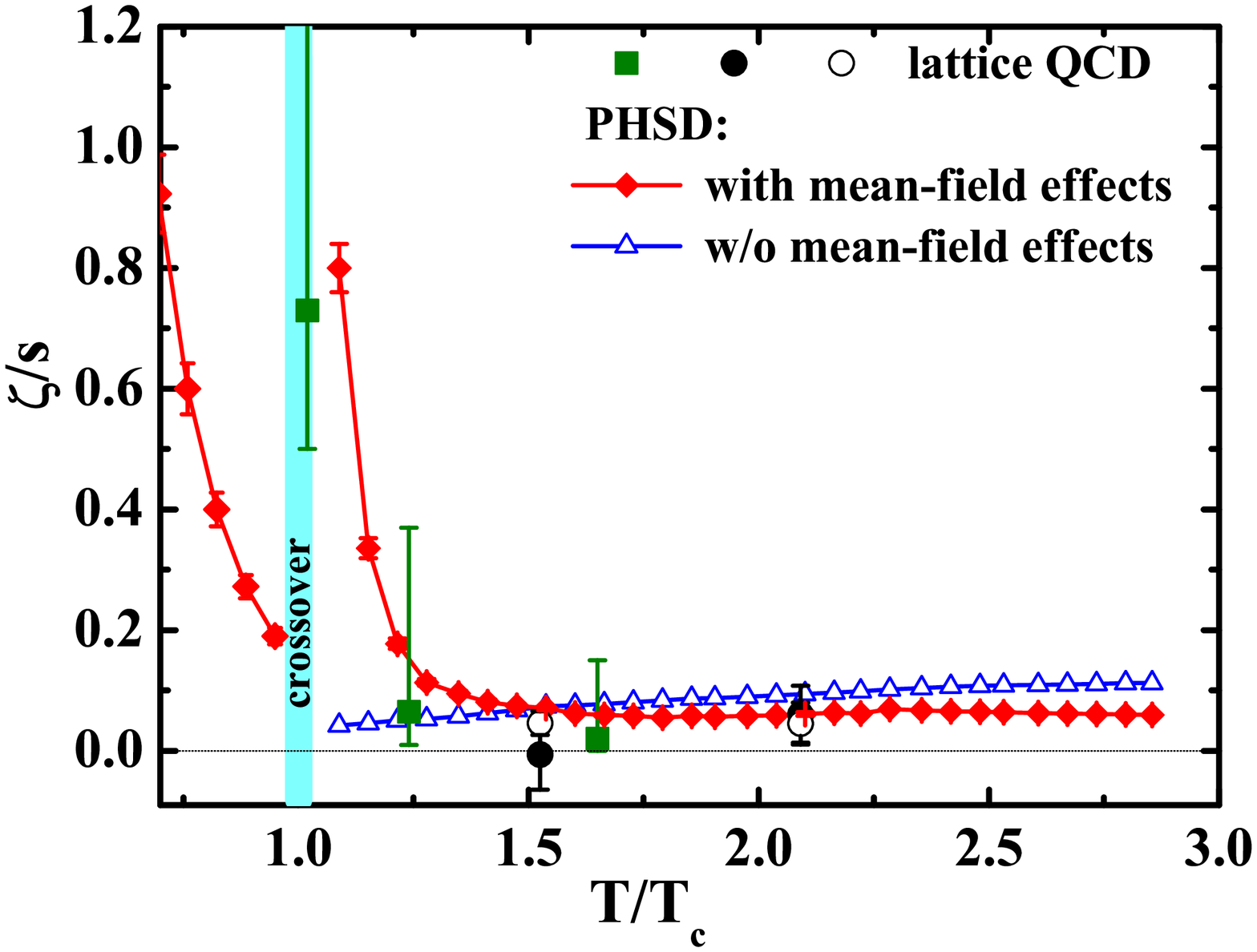}
\includegraphics[width=0.33\textwidth,clip]{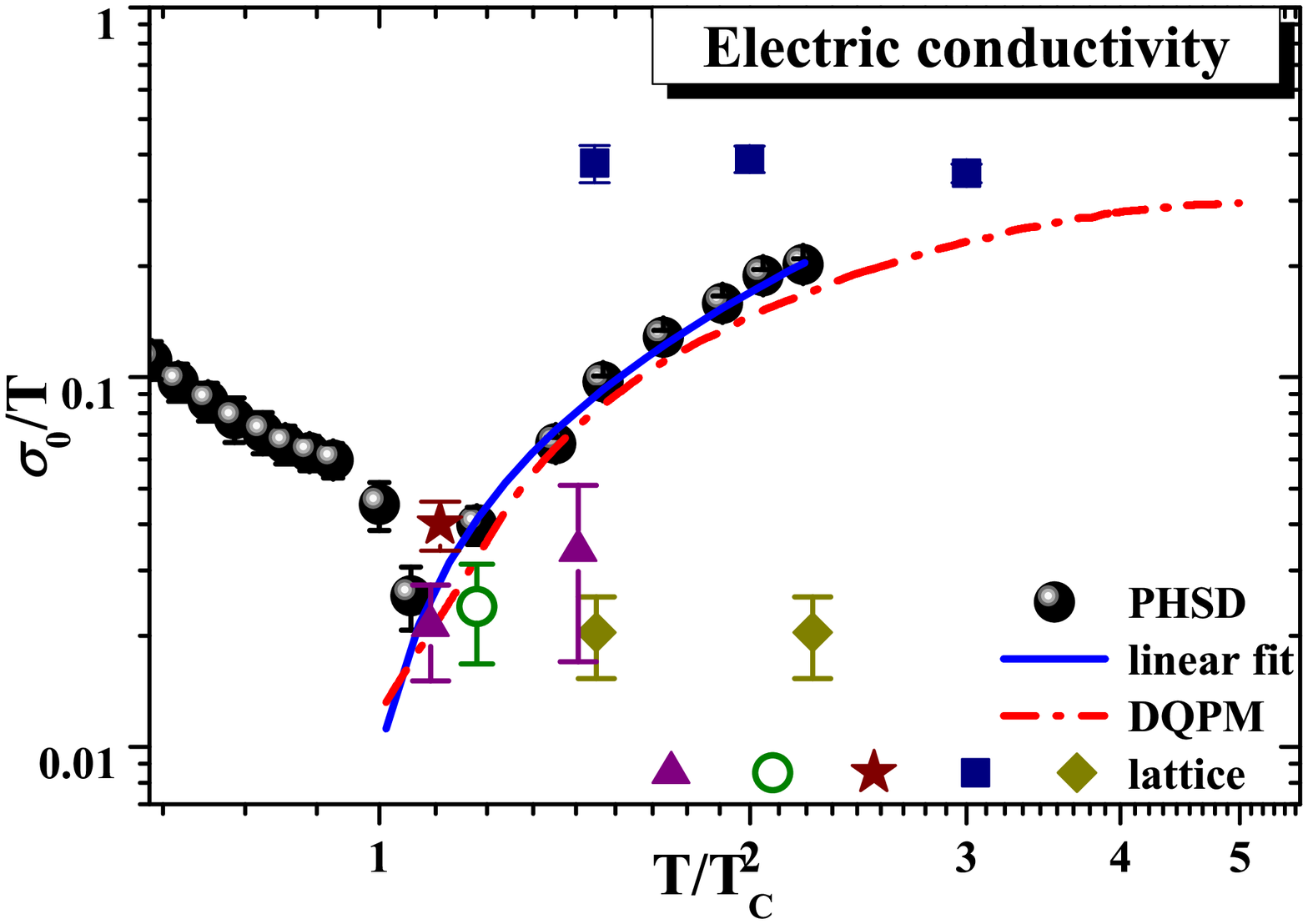}
\protect\caption{The PHSD results for the shear (upper part) and
bulk (middle part) viscosities of partonic and hadronic matter - as
well as the electric conductivity (lower part) - as a function of
the scaled temperature $T/T_c$.} \label{Fig_ON}
\end{figure*}

In Fig. \ref{Fig_ON} we present the PHSD results for the shear and
bulk viscosities of partonic and hadronic matter - as well as the
electric conductivity - as a function of the temperature $T/T_c$
($T_c$ = 158 MeV).
The
ratio of the shear viscosity to entropy density $\eta(T)/s(T)$ from
PHSD shows a minimum (with a value of about 0.1) close to the
critical temperature $T_c$, while it approaches the perturbative QCD
(pQCD) limit at higher temperatures in line with lattice QCD
results. For $T<T_c$, i.e. in the hadronic phase, the ratio $\eta/s$
rises fast with decreasing temperature due to a lower interaction
rate of the hadronic system and a significantly smaller number of
degrees-of-freedom.

The bulk viscosity $\zeta(T)$ -- evaluated in
the relaxation time approach -- is found to strongly depend on the
effects of mean fields (or potentials) in the partonic phase. We
find a significant rise of the ratio $\zeta(T)/s(T)$ in the
vicinity of the critical temperature $T_c$, which is also in
agreement with that from lQCD calculations. This rise has to be
attributed to mean-fields (or potential) effects that in PHSD are
encoded in the temperature dependence of the quasiparticle masses,
which is related to the infrared enhancement of the resummed
(effective) coupling $g(T)$.

We also find that the dimensionless
ratio of the electric conductivity over temperature $\sigma_0/T$
rises above $T_c$ approximately linearly with $T$ up to $T=2.5
T_c$, but approaches a constant above $5 T_c$, as expected
qualitatively from perturbative QCD (pQCD) (cf. Ref.
\cite{Cassing:2013iz} for  details). Our findings imply that
the QCD matter even at $T \! \approx \! T_c$ is a much better
electric conductor than $Cu$ or $Ag$ (at room temperature). We
note that our result for $\sigma_0/T$ close to $T_c$ is in agreement
with the most recent lQCD calculations which is important for
the photon emission rate from the QGP or
the hadronic system which is controlled by the electric conductivity.

\section{Summary}
Since the PHSD calculations have proven to describe
single-particle as well as collective observables  from relativistic
nucleus-nucleus collisions from lower SPS to top
RHIC energies \cite{Voka12}, the extracted transport coefficients $\eta(T)$,
$\zeta(T)$ and $\sigma_0(T)$ are compatible with experimental observations
in a wide energy density (temperature) range. Furthermore, the
qualitative and partly quantitative agreement with lQCD
transport coefficients is striking and should be further explored
in future.

\acknowledgements
This work in part has been supported by
the LOEWE center HIC for FAIR as well as DFG.

\end{document}